\documentstyle[prl,aps,multicol,exscale,epsf]{revtex}
\begin{document}

\onecolumn

\title{Structure, Structural Relaxation and Ion Diffusion in Sodium Disilicate 
       Melts} 

\author{A.~Meyer\,$^{1}$\cite{Meyer}, H.\ Schober\,$^{1,2}$, 
        D.\,B.\ Dingwell\,$^{3}$}
\address{$^{1}$\,Physik Department E\,13, Technische Universit\"at 
        M\"unchen, 85747 Garching, Germany}
\address{$^{2}$\,Institut Laue-Langevin, 38042 Grenoble, France}
\address{$^{3}$\,Institut f\"ur Mineralogie, Petrologie und Geochemie, 
        Universit\"at M\"unchen, 80333 M\"unchen, Germany}

\date{submitted to Europhysics Letters, January 7, 2002}

\maketitle

\begin{abstract}
We have investigated Na$_2$Si$_2$O$_5$ melts with inelastic neutron scattering 
at temperatures up to 1600\,K.
The Si-O network relaxes on a time scale of ns, whereas the Na ion 
relaxation dynamics are found on a time scale of 10\,ps. 
The elastic structure factor exhibits at $\simeq\!0.9$\,\AA$^{-1}$ an emerging 
prepeak which becomes more pronounced with increasing temperature. 
The prepeak is caused by the formation of sodium rich regions 
in the partially disrupted Si-O tetrahedral network.
\end{abstract}

\begin{multicols}{2}

Silicate melts have great relevance both in earth science and technology: 
The physical properties of magma (molten silicate rock) dominate many geological
processes and technological glasses are synthesized from the molten state.
These silicate melts, natural and technological, are 
multicomponent systems.
A considerable effort has been made to investigate a wide range of  
silicate melts 
with the general aim of linking structural with physical properties and to develop an 
atomic level understanding of their structure and dynamics \cite{revmin}.  
Nevertheless, experiments on the structure and microscopic dynamics in the viscous
melt well above the conventional glass transition temperature $T_g$ are rare.
Here, we present inelastic neutron scattering results on sodium
disilicate melts and show that at temperatures well above $T_g$
significant changes in the elastic structure factor occur 
that are caused by sodium rich regions 
in the partially disrupted Si-O tetrahedral network. 
The existence of sodium rich regions has a strong influence 
on the sodium ion transport
and may very well be the cause for the weak
temperature dependence of the viscosity at higher 
temperatures.
Our results on structure and dynamics can be rationalized using results
of recent molecular dynamics simulations \cite{HoKB99}. 

Alkali silicates are a simplified analog for most 
magmas and technological glasses. 
Compared with pure silica, with a caloric glass 
transition temperature $T_g$ at $\simeq\,1500$\,K,
the addition of Na$_2$O partially disrupts the SiO$_2$ network structure 
resulting in a halving of the $T_g$ at sodium disilicate. 
In addition, sodium disilicate has been
the subject of extensive molecular dynamics simulations \cite{HoKB99,OvSa98,VeGM92}
and thoroughly investigations \cite{MaSS83,KnDS94}.
Inelastic neutron scattering covers a dynamic range 
that gives insight not only into dynamics on microscopic time scales
but also into the medium range structure of silicate melts. 
In addition, the results can provide 
input for simulations used e.g.\ to model magma properties in volcanic
activity and are an experimental test of molecular dynamics simulations.

Sodium disilicate Na$_2$Si$_2$O$_5$ was synthesized from ultrapure 
Na$_2$CO$_3$ and SiO$_2$ powders. The mixture was melted
in a Pt crucible and stirred with a Pt$_{80}$Rh$_{20}$ spindle at superliquidus 
temperatures until it was homogeneous and bubble free.   
Differential scanning calorimetry yielded a glass transition
temperature at $\simeq\,$741\,K in agreement with the literature value at
$T_g\!\simeq\!737$\,K \cite{KnDS94}.
For the neutron scattering experiment the sample was encapsulated
in a Pt sample cell giving an annular sample geometry of 40\,mm 
in height, 22.5\,mm in diameter and a 1.25\,mm wall thickness.

We performed inelastic neutron scattering measurements
on the time-of-flight spectrometer IN\,6 at the 
Institut Laue-Langevin in Grenoble. 
An incident neutron wavelength of $\lambda\!=\!5.1\,\mbox{\AA}^{-1}$
yielded an energy resolution of $\delta E=93\,\mu\mbox{eV}$ (FWHM)
and an accessible wave number range at zero energy transfer
of $q=0.3-2.0\,\mbox{\AA}^{-1}$.
Spectra were measured in the glass at 400\,K 
and in the viscous melt at 800\,K and between 1200\,K and 1600\,K 
in steps of 100\,K.
During cooling measurements were repeated at each temperature.
No change between spectra obtained during heating and cooling 
could be detected. 
Additional high-energy resolution spectra at 300\,K, 1300\,K,
1400\,K, 1500\,K, and 1600\,K were obtained
on the new backscattering spectrometer \cite{MeDG01} 
of the NIST Center for Neutron Research in Gaithersburg
having a sub-$\mu$eV energy resolution and, in the set-up used, an  
accessible wave number range of $q=0.6-1.6\,\mbox{\AA}^{-1}$.

The raw data reduction consists of 
normalization to a vanadium standard, 
correction for self absorption and container scattering,
and interpolation to constant wave numbers $q$
in order to obtain the scattering law $S(q,\omega)$.
Further, $S(q,\omega)$ was symmetrized with respect to the energy transfer 
$\hbar\omega$ by means of the detailed balance factor.
Whereas scattering from the Si and O atoms is exclusively
coherent, Na scatters coherently and incoherently.
The incoherent scattering from Na reflects itself in
a flat background in the elastic structure factor 
$S(q,\omega\!=\!0)$.

Figure 1 displays the elastic structure factor of 
glassy and viscous sodium disilicate. 
Toward small $q$ the signal is dominated by incoherent 
contributions of the sodium atoms. 
The maximum around $\simeq\!1.7\,$\AA$^{-1}$ reflects the coherent 
scattering mainly of the disrupted tetrahedral Si-O network.
The intensity at $q$ values larger than $\simeq\!1.1$\,\AA$^{-1}$ is decreasing with
increasing temperature and $q$ which is well accounted for 
by the Debye--Waller factor. 
In contrast, at $q\!\simeq\!0.9$\,\AA$^{-1}$ a {\it pronounced prepeak} is emerging
with increasing temperature.
We note that the signal is independent on thermal history indicating
a process in thermoequilibrium as a cause for this change. 

Our $S(q,\omega\!=\!0)$ at 400\,K and 800\,K are consistent with
neutron diffraction results on sodium tetrasilicate Na$_2$Si$_4$O$_9$ 
\cite{ZoDK98} where at temperatures up to some 180\,K  
above $T_g$ spectra display only a little change in the structure factor. 
$^{29}$Si nuclear magnetic resonance (NMR) measurements on sodium disilicate 
\cite{MaYo97} in the metastable melt up to 876\,K report a 
continuous but small change in the number of O atoms surrounding a Si atom
with increasing temperature.
At higher temperatures motion of the atoms becomes too fast for conventional NMR techniques.
From wide angle X-ray diffraction on various sodium and potassium silicates
and sodium disilicate at 295\,K and 1273\,K in particular, 
previous workers \cite{WaSu77} came to the conclusion
that the structure of silicate melts is insensitive to temperature change. 
However, data quality especially in the $q$ range below $\simeq\,1.2$\,\AA$^{-1}$
is rather poor. 
The molecular dynamics simulations on sodium disilicate melts 
\cite{HoKB99} do indeed exhibit a shoulder in the static structure factor 
at $q\!=\!0.94$\,\AA$^{-1}$.
The molecular dynamics simulations \cite{HoKB99} link this prepeak 
to a characteristic length scale of regions where the 
network is disrupted and where the sodium concentration is enhanced.
 
Figure 2 shows high--energy resolution backscattering spectra of 
sodium disilicate melts at $q\!=\!1.6\,$\AA$^{-1}$. 
A measurement of the glassy sample at 300\,K represents the
instrumental resolution profile $R(q,\omega)$.
Above 1400\,K a small broadening of the signal can be resolved. 
Around  $q\!\simeq\!1.6\,$\AA$^{-1}$ 
the signal is dominated by the scattering of Si and O.
Structural relaxation of the partially disrupted tetrahedral Si-O network
leads to a quasielastic broadening in the spectra. 
A common feature of structural relaxation in glass forming liquids
is a stretching of correlation functions over a wider time range
than expected for exponential relaxation \cite{GoSj92,WoAn76}. 
The line is a fit with the Fourier transform of the Kohlrausch stretched 
exponential function convoluted with the instrumental resolution 
function:
\begin{equation}
R(q,\omega) \otimes \{A\,(\int\,dt\,e^{- i \omega t} 
\exp[-(t/\tau_q)^{\beta_q}]) + B\,\delta(\omega)\}.
\end{equation}
A best fit of expression (1) with $B\!=\!0$ gives a relaxation time of 4\,ns 
and a stretching exponent $\beta\!=\!0.75\pm\!0.1$.
$B\!=\!0$ demonstrates that the dynamics of all atoms contributing to the 
quasielastic signal are on the same time scale. 
Scattering on considerabely faster 
relaxation or vibration (ps time scale) results in a flat background.
At 1600\,K sodium disilicate has a viscosity of 4.5\,Pa\,s \cite{KnDS94}.
The relaxation time is consistent with the relaxation time estimated
from viscosity showing that the
data in Fig.\ 2 indeed display the network dynamics.
A fit with a stretched exponential function to the coherent intermediate 
scattering function
of Si and O correlations at 1.6\,\AA$^{-1}$ as obtained by the 
molecular dynamics simulations \cite{HoKB99,Hopriv}
gives $\beta\!=\!0.83$ in accordance with our result.

At temperatures up to $\simeq\!2\,\times\,T_g$ the diffusive 
dynamics of the alkali atoms in silicate melts is appreciably faster
than the structural relaxation \cite{DiWe90}.
Figure 3 displays the scattering law $S(q,\omega)$ of sodium disilicate
as obtained on the neutron time-of-flight spectrometer IN\,6. 
Well below $q\!=\!0.5$\,\AA$^{-1}$ the signal is dominated by the incoherent 
scattering on sodium.
Diffusive motion of sodium causes a broad quasielastic signal. 
The lines in Fig.\ 3 are fits with expression (1), where the 
$\delta$ function represents the elastic scattering contribution to the signal 
and $B$ its amplitude.
Best fits have been obtained with a stretching exponent $\beta$
within [0.7,0.8] 
in accordance with the MD simulations result
of $\beta=0.77$ for the incoherent intermediate Na scattering function 
\cite{HoKB99,Hopriv}.
$\beta < 1$ indicates that Na diffusion is not a simple activated 
process.
Resulting relaxation times (Fig.\ 3) show a small change with temperature 
with $\tau\!\simeq\!29$\,ps at 1200\,K and $\tau\!\simeq\!13$\,ps at 1600\,K. 

Compared to other multicomponent glass forming liquids 
investigated so far the changes in the elastic structure factor of 
sodium disilicate melts are unexpected.
Viscous metallic ZrTiNiCuBe melts show a similar
temperature dependence of viscosity \cite{MaWB99} 
as sodium disilicate when plotted versus $T_g/T$
at temperatures well below the liquidus.
In addition, diffusive dynamics of various atoms, that are smaller than
the large structural Zr atoms, are appreciably faster than structural 
relaxation and also decoupled from viscous flow \cite{EhHR98}.
Toward higher temperatures the temperature dependence of the viscosities
of ZrTiNiCuBe and sodium disilicate deviate \cite{KnDS94} which may very well
be the result of the sodium rich regions in sodium disilicate:
The formation of sodium rich regions leaves regions with a lower
sodium concentration and hence with a less disrupted Si-O network
leading to a higher viscosity of the entire system.

Besides a non-exponential structural relaxation viscous metallic melts  
exhibit a fast relaxation process \cite{metglas} 
which can be visualized as a rattling of the atoms in the transient cages 
formed by their neighbors. 
This -- so called -- fast $\beta$ relaxation, 
that prepares structural relaxation responsible 
for viscous flow, has been predicted by the
mode coupling theory (MCT) of the liquid to glass transition \cite{GoSj92}
and has been found in many experiments and simulations on a variety of 
other glass forming systems \cite{Goe99}.
Molecular dynamics simulations on sodium disilicate 
show that the predicted factorization property for Si and O correlations 
holds in the fast $\beta$ relaxation regime \cite{Hopriv}.
The signal in the fast $\beta$ relaxation regime, typically 
around 1\,meV, is in sodium disilicate governed by the 
incoherent contributions of the diffusing sodium atoms. 
However, since the sodium disilicate spectra do not show an indication 
for an additional fast process, its amplitude has to be small compared
to the quasielastic line originating from the sodium diffusion. 
This compares well to results of a neutron scattering experiment 
on another network forming melt: In liquid GeO$_2$
a fast MCT $\beta$ relaxation process could not be detected 
up to $T\!=\!2\times T_g$ \cite{MeSN01}.

A recent molecular dynamics simulations study on sodium tetrasilicate
\cite{JuKJ01} suggests the existence of sodium rich channels having a distance of 
about 6-8\,\AA . 
The prepeak at 0.9\,\AA$^{-1}$ corresponds to the distance between
these channels.
A similar behavior has been found in the MD simulations on sodium 
disilicate \cite{Hopriv}.
The microscopic picture given by the molecular dynamics simulations 
\cite{HoKB99,Hopriv,JuKJ01}
of regions that exhibit a higher sodium concentration is in agreement 
with our results. 

In conclusion, 
we have investigated structure and dynamics in sodium disilicate melts with inelastic
neutron scattering. 
The viscous melt well above $T_g$
exhibits changes in the elastic structure factor: With increasing temperature a
pronounced prepeak emerges at $\simeq\!0.9$\,\AA$^{-1}$ caused by
sodium rich regions in the disrupted Si-O network.
High resolution quasielastic neutron scattering shows structural relaxation of 
the partially disrupted Si-O tetrahedral network to be  
on a nanosecond time scale at 1600\,K. 
The structural relaxation is non-exponential
with a stretching exponent $\beta\!\simeq0.75$.
Inelastic neutron scattering reveals diffusive dynamics of the sodium ions 
on a 10\,ps time scale. Even in the equilibrium melt 
the Sodium ions diffuse in a relative immobile Si-O matrix.

We thank W.\ Petry for his support and J.\ Horbach and W.\ Kob for fruitful 
discussions. We acknowledge the support of the National Institute of Standards and Technology, 
U.\ S.\ Department of Commerce, in providing the neutron backscattering spectrometer
used in this work and we thank D.\ Neumann and R.\,M.\ Dimeo 
for their support.


\newpage

\narrowtext

\begin{figure}[t]
\epsfxsize=7.6cm
\leavevmode \epsffile{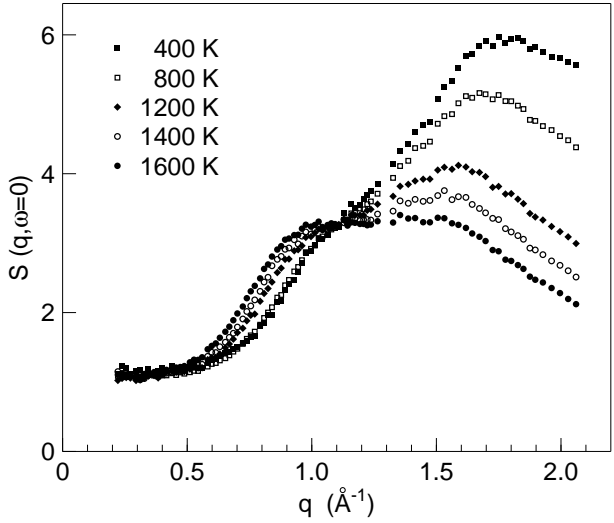}
\caption{Elastic structure factor $S(q,\omega\!=\!0)$ of 
sodium disilicate: Toward small $q$ the signal is dominated by the incoherent 
contributions of the sodium atoms. 
The maximum around $\simeq\!1.7\,$\AA$^{-1}$ reflects the scattering of the 
partially disrupted tetrahedral SiO$_4$ network.
With increasing temperature a pronounced shoulder is emerging at 
$q\!\simeq\!0.9$\,\AA$^{-1}$. The decrease in intensity at 
$q$ values larger than $\simeq\!1.1$\,\AA$^{-1}$ in turn 
mainly reflects the Debye--Waller factor. 
}
\end{figure}

\begin{figure}
\epsfxsize=7.6cm
\leavevmode \epsffile{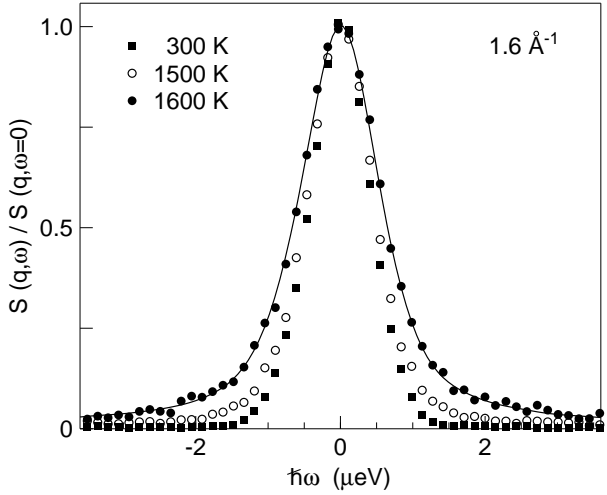}
\caption{High--energy resolution backscattering spectra of 
a sodium disilicate melt. Around  $q\!\simeq\!1.6\,$\AA$^{-1}$ 
the signal is dominated by the coherent scattering of Si and O.
At 1600\,K structural relaxation of the Si-O network
is on a ns--time scale causing a broadening of the quasielastic line. 
The line represents a fit 
(1) giving a relaxation time of $\simeq 4$\,ns and a stretching
exponent $\beta\!=\!0.75\pm\!0.1$.
}
\end{figure}

\begin{figure}
\epsfxsize=7.6cm
\leavevmode \epsffile{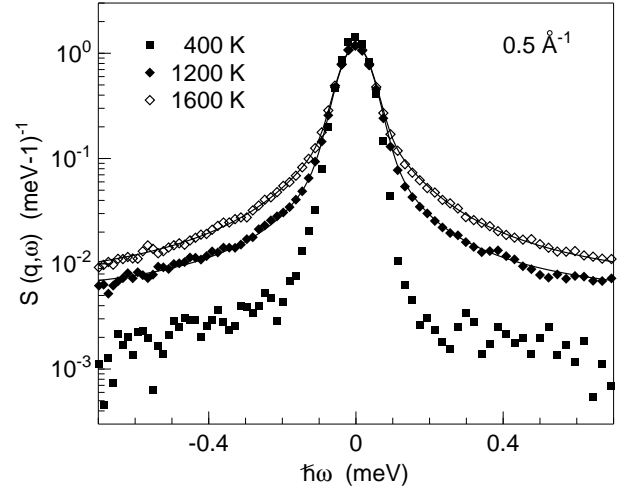}
\caption{Scattering law $S(q,\omega)$ at $q\!=\!0.5\,$\AA$^{-1}$
measured on the neutron time-of-flight spectrometer IN\,6. 
Diffusion of sodium ions results in a broad quasielastic signal
that is described by (1) with a stretching exponent $\beta$
within [0.7,0.8] and relaxation times $\tau$ at some 29\,ps at
1200\,K and some 13\,ps at 1600\,K. 
}
\end{figure}

\end{multicols}

\end{document}